\documentclass{moriond}

\usepackage{amsmath}
\usepackage{wrapfig}

\def\Journal#1#2#3#4{{#1} {\bf #2}, #3 (#4)}

\def\PRD{{\em Phys. Rev.} D}

\def\PRD{{\em Phys. Rev.} D}
\def\MNRAS{{\em Mon. Not. Roy. Astron. Soc.}}
\def\ApJL{{\em Astrophys. J. Lett.}}
\def\AandA{{\em Astron. Astrophys.}}
\def\PASJ{{\em Publ. Astron. Soc. Japan}}
\def\PASP{{\em Publ. Astron. Soc. Pac.}}

\def\be{\begin{equation}}
\def\ee{\end{equation}}
\def\bea{\begin{eqnarray}}
\def\eea{\end{eqnarray}}

\newcommand{\arXiv}[1]{\href{https://arxiv.org/abs/#1}{arXiv:#1}}

\begin{document}
\vspace*{3cm}
\title{The {\sc Lemaître} project: An Independent Probe of the $w-w_a$ Tension }

\author{Nicolas Regnault$^1$, on behalf of the {\sc Lemaître} Project}

\address{$^1$ Sorbonne Université, CNRS, Université Paris-Cité, LPNHE, 75252 Paris Cedex 05, France.}

\maketitle

\abstracts{Recent BAO, CMB, and Type Ia supernova constraints on the
  dark-energy equation of state show a tension with the cosmological-constant
  model at the heart of $\Lambda$CDM, partly driven by low-redshift supernova
  measurements. We present the {\sc Lema\^itre} project, a new measurement of
  the dark energy equation of state, based on a distinct SN~Ia dataset (2800
  from ZTF, 170 from SNLS and 300 from HSC), a dedicated calibration program,
  and a new, independent analysis pipeline.}

\section{Introduction: the {\sc Lemaître} project}

Recent constraints from DESI BAO, combined with CMB anisotropies and Type Ia
supernovae favor a dynamical dark-energy component over a cosmological constant.
In the CPL parametrization of the dark energy equation of state, $w(a)=w_0 + w_a
(1-a)$, the latest multi-probe combinations point towards $w_0>-1$ and
$w_a<0$~\cite{DESI2025,Union3,DES5yr}, indicating an equation of state that lies
in the quintessence regime today, but crosses the phantom divide $w= -1$ at
intermediate redshift. Such a crossing cannot be realized by a single minimally
coupled canonical scalar field~\cite{Vikman2005}, and therefore points, if
confirmed, to a more complex dark-energy sector~\cite{Hu2005}.

This tension has been reduced however, by the recalibration of the DES-5y
supernova sample~\cite{popovic2025} and by recent updates to the Union
host-galaxy environment modelling~\cite{Union31}, yet a $\sim 3 \sigma$
departure from $\Lambda$CDM persists in the DESI+CMB+SN combination. Since the
residual tension is very sensitive to supernova calibration and possible cross-survey
systematics, an independent test of the supernova contribution is essential.

The \textsc{Lema\^itre} project\,\footnote{Latest Extended Mapping of Acceleration with an Independent Trove of Redshifted Explosions} addresses this need by providing an independent
measurement of the dark-energy equation of state based on three pillars: (1) a
distinct supernova dataset, unused in previous cosmological analyses (2) a
metrology and flux-calibration program designed to control measurement
systematics at the per-mil level (3) a new, lightweight cosmological inference
pipeline, therefore achieving a new and independent measurement of the
dark-energy equation of state.

\section{Requirements}
\label{sec:requirements}

\begin{wrapfigure}{r}{0.5\textwidth}
  \centering
  \vspace{-10pt}
  \includegraphics[width=0.5\textwidth]{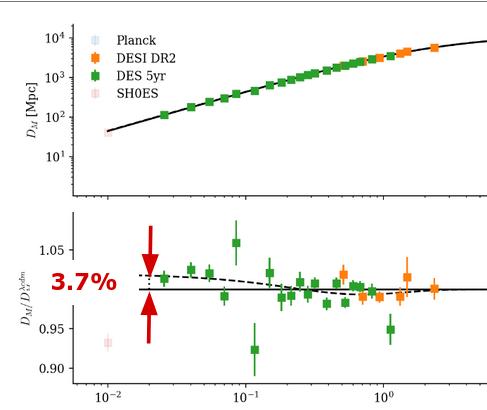}
  \caption{The impact of the $w-w_a$ tension in the combined SN/BAO Hubble
    diagram. }
  \label{fig:hd}
  \vspace{-10pt}
\end{wrapfigure}

The $w-w_a$ tension translates into a $\simeq 0.04$-mag distance modulus shift
with respect to $\Lambda$CDM, at $z \leq 0.1$ (Figure \ref{fig:hd}).
This is precisely the redshift range where Type Ia supernovae provide the only
direct handle on the distance scale. This is also the range where current
supernova samples remain statistically limited and heterogeneous.

Addressing this difference requires an accuracy which is a small fraction of the
offset measured in the Hubble diagram. A natural target is a 0.004-mag
uncertainty on the mean distance modulus in the low-redshift bin, corresponding
to roughly 10\% of the effect. Achieving this goal requires both statistical and
systematic uncertainties to be controlled below this level.

\paragraph{Statistical uncertainties} In the Tripp standardization
scheme~\cite{tripp1998}, the object-to-object dispersion of Type Ia supernovae
is typically $\sigma_{\mathrm{SN}} \sim 0.15$ mag.
As a consequence, at least 1500 supernovae are required in a given redshift bin to reach
a $0.004$~mag statistical uncertainty on the mean distance modulus: $\sigma_z \approx \frac{0.15}{\sqrt{1500}} \approx 0.0039\ \mathrm{mag}$.
Since the cosmological signal is measured differentially between low- and
high-redshift samples, a high-redshift sample of comparable statistical power is
also required.

\paragraph{Systematic uncertainties}
Recent supernova cosmology analyses typically track more than twenty distinct
sources of systematic uncertainty~\cite{vincenzi2024,rubin2025}. While
astrophysical systematics can be constrained by increasing the size and
diversity of the sample, instrumental systematics such as flux calibration and
bandpass uncertainties are coherent within each survey and nearly degenerate
with the cosmological signal.

Calibration requirements can be summarized in terms of flux zero-point
uncertainties, $\sigma_{\mathrm{zp}}$, and effective-bandpass mean wavelength uncertainties,
$\sigma_\lambda$. A crude propagation leads to the following relations between the distance modulus and these quantities:
$\sigma_\mu \simeq 3.6\sigma_{\mathrm{zp}}$ and
$\sigma_\mu \simeq 11\sigma_\lambda~{\rm mmag\,nm^{-1}}$. Requiring a
4-mmag accuracy on distance moduli therefore implies
$\sigma_{\mathrm{zp}}\simeq 0.001$ mag and
$\sigma_\lambda\simeq 3.6~\text{\AA}$.

\section{Dataset}
\label{sec:dataset}

The {\sc Lema\^itre} dataset consists of approximately 3300 Type Ia supernovae
covering the redshift range $0.02 < z < 1.1$, selected from three homogeneous
still unpublished samples.

\paragraph{ZTF} The low-redshift sample is based on the first three years of operation
of the Zwicky Transient Facility (ZTF)~\cite{bellm2019}. ZTF uses the robotic
48-inch Oschin Schmidt telescope (1.2-m) at Palomar Observatory, equipped with a
dedicated 47 deg$^2$ camera, to carry out a high-cadence survey of the northern
sky in three optical bands ($g$, $r$, and $i$). The ZTF observing system is
complemented by a dedicated low-resolution integral-field spectrograph,
operating on the Palomar 60-inch telescope (P60), which provides spectroscopic
classifications for ZTF transients brighter than approximately 19-mag. ZTF has
proved to be an exceptionally efficient SN~Ia discovery machine: over seven
years of operation (2018--2025), it has produced a homogeneous sample of more
than 10,000 spectroscopically classified Type Ia supernovae. The first three
years of ZTF observations form the basis of the ZTF second data release
(DR2)~\cite{rigault2024}. After quality cuts, this sample contains about 2800 cosmology-grade supernovae.

For the {\sc Lema\^itre} project, this sample is being reprocessed and
recalibrated in order to meet the accuracy requirements outlined above. A large
calibration catalog covering the entire northern sky was built and anchored on
the CALSPEC primary flux standard network with a precision of 1-mmag in all
ZTF-bands~\cite{VanDenAbeele2026a,Racine2026}. The ZTF bandpasses were measured
{\em in-situ} using a Traveling Collimated Beam projector
(TCBP)~\cite{VanDenAbeele2026b,Sommer2024}: a calibrated source that injects a
monochromatic pencil beam in the telescope optics. This measurement campaign
similarly enabled a remapping of the non-uniformities of the ZTF bandpasses and
yielded a sensor-dependent bandpass model with uncertainties at the $\sim
1\ \text{\AA}$-level~\cite{VanDenAbeele2026a}, which meets the accuracy goal
discussed above.

\paragraph{SNLS} The ZTF sample is complemented at higher
redshift by 172 Type Ia supernovae observed during the final two years of the
SNLS survey, which had not previously been used in cosmological analyses.

A dedicated calibration program was carried out during engineering time to
anchor the supernova observations directly to primary flux calibrators. The
precision of this anchoring is of 2 to 5 mmag, depending on the band. In
addition, the decommissioned MegaCam filters were remeasured on a dedicated
bench built at the Laboratoire des Matériaux Avancés (LMA, Lyon)~\cite{Sassolas2018}.
This measurement campaign allowed to remap the sizeable non-uniformities of
each filter, and to determine focal-plane position-dependent filters with an
accuracy of 2 to 4~\AA, depending on the band.

\paragraph{HSC} The very-high-redshift end of the dataset is provided by
a large sample of about 300 Type Ia supernovae discovered and followed up with
the 1.8 deg$^2$ Hyper Suprime-Cam instrument mounted on the 8.2-m Subaru
telescope. This Subaru transient search is based on a recadencing of the Deep
and UltraDeep layers of the Hyper Suprime-Cam Subaru Strategic Program. The
resulting sample covers the redshift range $0.6 < z < 1.1$, with particularly
strong constraining power at $0.8 < z < 1.1$. This is made possible by the
deep-depletion, red-sensitive HSC CCDs, which provide high signal-to-noise
observations in the $z$ and $Y$ bands.

The HSC bandpasses were measured on a dedicated bench at NAOJ, using a
monochromatic beam designed to reproduce the geometry of the science
beam~\cite{Kawanomoto2018}. Unlike the ZTF and SNLS samples, the Subaru/HSC sample
is not tied directly to CALSPEC standards, since these stars are too bright to
be observed with HSC. However, the three {\sc Lema\^itre} samples have
overlapping footprints and therefore share calibration stars. This provides a
direct route to cross-calibrating the Subaru, SNLS, and ZTF samples within a
common photometric system. Moreover, since both the SNLS and ZTF calibration
chains include direct observations of CALSPEC standards, the combined
calibration network provides two independent anchors to the primary flux scale.

\paragraph{} Overall, the combined sample has a statistical power comparable to DES~5Y,
with substantially stronger low-redshift coverage.

\section{Photometry pipeline}
\label{sec:photometry_pipeline}

To avoid redshift-dependent biases from survey-dependent photometry choices, all
three samples are processed with the same scene-modeling photometry software and
calibration pipeline.

Supernova fluxes are measured using a scene-modeling photometry approach, in
which the supernova plus host-galaxy scene is forward-modeled at the pixel
level. For image (i), the model can be written schematically as
\begin{equation}
\phi_{i,p} = f_i\,\psi_{i,p} + \left(K_i \otimes G\right)_p ,
\end{equation}
where $f_i$ is the supernova flux in image $i$, $\psi_{i,p}$ is the PSF of that
image evaluated at pixel $p$, $K_i$ is the convolution kernel that maps the
reference-frame galaxy model to image $i$, and $G$ is the host-galaxy background
model, typically represented as a grid of free pixels in the frame of the
best-seeing image. After the fit, the supernova fluxes $f_i$ are retained as the
quantities of interest, while the nuisance parameters describing the galaxy
background are marginalized over.

This estimator is close to statistically optimal for low-SNR high-redshift
measurements; more importantly it can be applied to field stars, allowing
photometry biases to cancel between scene modeling and calibration measurements
-- up to residual non-linearity effects.

\paragraph{Sensor effects} At the targeted accuracy, sensor-induced PSF
distortions require explicit correction. This includes the brighter-fatter
effect in the deep-depletion HSC sensors~\cite{Astier2023}, and the ZTF ``pocket
effect'', a flux-dependent PSF skewness induced by charge trapping near the
serial register. Dedicated pixel-level corrections have been developed and
applied to the survey images.

\paragraph{} The scene-modeling light curves of the supernovae and field stars
are injected into a unique calibration pipeline, which aligns the three surveys
on the same flux scale, namely the HST white dwarf flux scale, implemented by
the CALSPEC system~\cite{Bohlin2014}. The resulting relative calibration
precision $\sigma_{zp}$ is of 1~mmag for ZTF and 2-5 mmag for SNLS and HSC. The
supernova light curves are subsequently blinded before entering the cosmological
analysis.

\section{Cosmology Inference pipeline}
\label{sec:inference_pipeline}

To ensure the independence of the cosmological results, we are developing a new
cosmology inference chain. This effort is driven not only by the need for an
independent implementation, but also by the desire to simplify the analysis as
much as possible, making the full chain lightweight, transparent, and testable
on large ensembles of simulations.

The resulting pipeline is organized into three main components. The first,
called NaCl, builds a spectrophotometric model of the Type Ia supernova
population from the photometric and spectroscopic follow-up of the cosmology
sample. The second, EDRIS~\cite{Kuhn2025}, converts the fitted $(x_0,x_1,c)$
triplets into standardized luminosity distances, while simultaneously
determining the survey selection functions and correcting for survey-dependent
Malmquist biases. The final component performs the cosmological inference
itself, combining the binned distance constraints produced by EDRIS with
external cosmological priors from Planck and DESI. This step relies in
particular on {\tt cosmologix}~\cite{Betoule2026}, a fast cosmology-inference
code written in JAX.

The three components are integrated into a single end-to-end inference chain,
which is currently being validated through a sequence of data challenges based
on simulations of increasing complexity.

\section{Conclusion}

The {\sc Lema\^itre} project has completed its metrology program and is now
releasing blinded light curves internally. Deblinding will proceed once the
end-to-end inference chain has passed the ongoing Data Challenges.

\section*{Acknowledgements}

\footnotesize{Support for this project was provided by the Agence Nationale de
  la Recherche of the French government under grants ANR-21-CE31-0016-03
  (ZTF-II) and ANR-23-CE31-0007 (CARL), and by the European Research Council
  (ERC) under grant 759194-USNAC. We gratefully acknowledge the ZTF telescope
  staff at Caltech and Palomar Observatory, the CFHT staff on Maunakea, and the
  NAOJ/Subaru staff for their assistance with engineering observations and
  dedicated measurement campaigns. }

\end{document}